\documentclass[aps,tightenlines,twocolumn,nofootinbib,superscriptaddress]{revtex4-1}

\usepackage{graphicx}
\usepackage{subfigure}
\usepackage{epstopdf}
\usepackage{dcolumn}
\usepackage{bm}
\usepackage{xcolor}%

\usepackage{booktabs}
\usepackage{makecell}
\usepackage{threeparttable}
\usepackage{pifont}

\newcommand{\Fig}{Fig.~}

\begin{document}
\title{EBS and in-beam $\gamma$-ray investigation of FCVA-prepared Cr$_2$O$_3$ targets for low-energy $^{16}$O+$^{16}$O fusion experiments} 

\author{C.~Wen}
\affiliation{Key Laboratory of Radiation Physics and Technology of the Ministry of Education, Institute of Nuclear Science and Technology, Sichuan University, Chengdu 610064, China}

\author{X.~Chen}
\affiliation{Key Laboratory of Beam Technology of Ministry of Education, School of Physics and Astronomy, Beijing Normal University, Beijing, 100875, China}

\author{L.~Wang}
\affiliation{Key Laboratory of Beam Technology of Ministry of Education, School of Physics and Astronomy, Beijing Normal University, Beijing, 100875, China}

\author{Z.~An}
\affiliation{Key Laboratory of Radiation Physics and Technology of the Ministry of Education, Institute of Nuclear Science and Technology, Sichuan University, Chengdu 610064, China}

\author{F.~Bai}
\affiliation{Key Laboratory of Radiation Physics and Technology of the Ministry of Education, Institute of Nuclear Science and Technology, Sichuan University, Chengdu 610064, China}

\author{Y.~Chen}
\affiliation{Key Laboratory of Radiation Physics and Technology of the Ministry of Education, Institute of Nuclear Science and Technology, Sichuan University, Chengdu 610064, China}

\author{X.~Fang}
\affiliation{Sino-French Institute of Nuclear Engineering and Technology, Sun Yat-sen University, Zhuhai, Guangdong 519082, China}

\author{Y.X.~Fan}
\affiliation{Key Laboratory of Radiation Physics and Technology of the Ministry of Education, Institute of Nuclear Science and Technology, Sichuan University, Chengdu 610064, China}

\author{B.S.~Gao}
\affiliation{State Key Laboratory of Heavy Ion Science and Technology, Institute of Modern Physics, Chinese Academy of Sciences, Lanzhou 730000, China}
\affiliation{School of Nuclear Science and Technology, University of Chinese Academy of Sciences, Beijing 100049, China}
\affiliation{Joint Department for Nuclear Physics, Lanzhou University and Institute of Modern Physics, Chinese Academy of Sciences, Lanzhou 730000, China}

\author{Z.Y.~Guo}
\affiliation{Key Laboratory of Radiation Physics and Technology of the Ministry of Education, Institute of Nuclear Science and Technology, Sichuan University, Chengdu 610064, China}

\author{J.F.~Han}
\affiliation{Key Laboratory of Radiation Physics and Technology of the Ministry of Education, Institute of Nuclear Science and Technology, Sichuan University, Chengdu 610064, China}

\author{H.T.~Hu}
\affiliation{Key Laboratory of Beam Technology of Ministry of Education, School of Physics and Astronomy, Beijing Normal University, Beijing, 100875, China}

\author{W.P.~Lin}
\email{linwp1204@scu.edu.cn}
\affiliation{Key Laboratory of Radiation Physics and Technology of the Ministry of Education, Institute of Nuclear Science and Technology, Sichuan University, Chengdu 610064, China}

\author{B.~Liao}
\affiliation{Key Laboratory of Beam Technology of Ministry of Education, School of Physics and Astronomy, Beijing Normal University, Beijing, 100875, China}

\author{S.~Lin}
\affiliation{Key Laboratory of Beam Technology of Ministry of Education, School of Physics and Astronomy, Beijing Normal University, Beijing, 100875, China}

\author{G.~Liu}
\affiliation{Key Laboratory of Radiation Physics and Technology of the Ministry of Education, Institute of Nuclear Science and Technology, Sichuan University, Chengdu 610064, China}

\author{X.Q.~Liu}
\affiliation{Key Laboratory of Radiation Physics and Technology of the Ministry of Education, Institute of Nuclear Science and Technology, Sichuan University, Chengdu 610064, China}

\author{P.P.~Ren}
\affiliation{Key Laboratory of Radiation Physics and Technology of the Ministry of Education, Institute of Nuclear Science and Technology, Sichuan University, Chengdu 610064, China}

\author{J.~Su}
\email{sujun@bnu.edu.cn}
\affiliation{Key Laboratory of Beam Technology of Ministry of Education, School of Physics and Astronomy, Beijing Normal University, Beijing, 100875, China}

\author{J.H.~Tan}
\affiliation{Key Laboratory of Radiation Physics and Technology of the Ministry of Education, Institute of Nuclear Science and Technology, Sichuan University, Chengdu 610064, China}

\author{X.D.~Tang}
\affiliation{State Key Laboratory of Heavy Ion Science and Technology, Institute of Modern Physics, Chinese Academy of Sciences, Lanzhou 730000, China}
\affiliation{School of Nuclear Science and Technology, University of Chinese Academy of Sciences, Beijing 100049, China}
\affiliation{Joint Department for Nuclear Physics, Lanzhou University and Institute of Modern Physics, Chinese Academy of Sciences, Lanzhou 730000, China}

\author{P.~Wang}
\affiliation{Key Laboratory of Radiation Physics and Technology of the Ministry of Education, Institute of Nuclear Science and Technology, Sichuan University, Chengdu 610064, China}

\author{S.~Wang}
\affiliation{Key Laboratory of Radiation Physics and Technology of the Ministry of Education, Institute of Nuclear Science and Technology, Sichuan University, Chengdu 610064, China}

\author{D.H.~Xie}
\affiliation{Key Laboratory of Radiation Physics and Technology of the Ministry of Education, Institute of Nuclear Science and Technology, Sichuan University, Chengdu 610064, China}

\author{N.T.~Zhang}
\affiliation{State Key Laboratory of Heavy Ion Science and Technology, Institute of Modern Physics, Chinese Academy of Sciences, Lanzhou 730000, China}

\date{\today}

\begin{abstract}
The Cr$_2$O$_3$ solid target for low-energy $^{16}$O+$^{16}$O fusion experiments was fabricated using filtered cathodic vacuum arc (FCVA) deposition. Its composition, oxygen areal density and impurity content were characterized by elastic backscattering spectrometry (EBS), and the impurity-induced background contributions were investigated by in-beam $\gamma$-ray spectroscopy. EBS results indicate that the Cr$_2$O$_3$ film exhibits good stoichiometry and uniformity, with $^{16}$O areal densities ranging from $(3.24$\textendash$3.25)\times10^{17}$~atoms/cm$^2$.
The EBS analysis reveals a carbon atomic fraction of approximately 1.25\textendash1.29\% in the Cr$_2$O$_3$ layer, while a large amount of carbon impurities are also identified on the surface of Cr substrate.
In-beam $\gamma$-ray spectra reveal prominent transitions associated with $^{27}$Al and $^{24}$Mg at 844, 1015, and 1369~keV, mainly originating from $^{12}$C+$^{16}$O fusion reactions induced by carbon impurities under $^{16}$O irradiation.
Meanwhile, characteristic $\gamma$-rays emissions from evaporation channels of the $^{16}$O+$^{16}$O reaction, including $^{31}$S, $^{31}$P, and $^{28}$Si, were also observed and can be used to extract the $^{16}$O+$^{16}$O fusion cross sections. 
This work provides an experimental basis for the development of high-purity oxide targets and the optimization of target configurations for future low-background $^{16}$O+$^{16}$O fusion cross section measurements.
\end{abstract}


\maketitle

\section{Introduction}
The $^{16}$O+$^{16}$O fusion reaction is one of the key heavy-ion fusion processes in the late evolution stage of massive stars. It plays an important role in oxygen burning and in the production of intermediate-mass nuclei in stellar environments\cite{gasquesImplicationsLowenergyFusion2007,yakovlevFusionReactionsMulticomponent2006}. In particular, the fusion cross sections and astrophysical S factors at sub-barrier energies strongly affect the ignition conditions, energy generation, and elemental abundances in stellar evolution and explosive astrophysical environments\cite{dumontAdvancedEvolutionMassive2025}. Therefore, precise measurements of the low-energy $^{16}$O+$^{16}$O fusion reaction are important for understanding stellar nucleosynthesis and improving astrophysical reaction models.
During the past decades, this reaction has been studied both theoretically\cite{diaz-torresEffectsNuclearMolecular2007, two} and experimentally\cite{duarteMeasurementFusionCross2015, hulkeComparisonFusionReactions1980, kuronenCrossSection16O+16O1987, spinkaExperimentalDeterminationTotal1974, thomasSubbarrierFusionOxygen1986, wuFusionElasticScattering1984}. For a typical oxygen-burning temperature of $T \approx 2.2$~GK, the corresponding Gamow window is located in the center-of-mass energy range of $5.3~\mathrm{MeV}<E_{\mathrm{c.m.}}<7.9~\mathrm{MeV}$, where the fusion cross sections are about $10^{-9}$--$10^{-3}$~barn. However, the lowest experimental energy reached so far is only $E_{\mathrm{c.m.}}=6.74$~MeV, which does not fully cover the astrophysical energy region. Direct measurements of the $^{16}$O+$^{16}$O fusion reaction in the lower-energy part of the astrophysical energy region remain very challenging because the fusion cross sections decrease rapidly with decreasing energy.

In low-energy heavy-ion fusion experiments, target properties are one of the main sources of systematic uncertainty. The target thickness, stoichiometric composition, and impurity content must be well known in order to determine the effective reaction energy, beam energy loss, and reaction yield normalization accurately. Impurity elements can also produce parasitic nuclear reactions under oxygen beam bombardment and generate extra $\gamma$-ray backgrounds, which may affect the extraction of the fusion cross sections. 
The $^{16}$O+$^{16}$O fusion cross section is usually determined by measuring the characteristic $\gamma$ rays from evaporation residues, such as $^{24}$Mg, $^{27}$Al, $^{28}$Si, etc. Carbon contamination is particularly critical because some of the same residual nuclei, including $^{24}$Mg and $^{27}$Al, can also be produced through the $^{12}$C+$^{16}$O fusion reaction, whose cross section is 2--3 orders of magnitude larger than that of the $^{16}$O+$^{16}$O fusion reaction in the low-energy region. Even a small amount of carbon contamination can therefore produce strong $\gamma$-ray backgrounds under intense oxygen irradiation, leading to possible overestimation of the fusion yield.
Therefore, reliable target characterization and impurity evaluation are essential for high-precision $^{16}$O+$^{16}$O fusion measurements.

At present, a number of mature methods have been developed for the preparation of solid oxide targets, including ion implantation~\cite{silvaComparativeAnalysisAnodized2014,silvaProductionThinTargets2017}, anodic oxidation~\cite{amsel1978precisiona,lunacollaborationPreparationCharacterisationIsotopically2012}, and magnetron sputtering~\cite{li2017preparation, li2024preparation}. In this work, we propose a new approach based on filtered cathodic vacuum arc (FCVA)\cite{tay2006review} technology for the fabrication of chromium oxide (Cr$_2$O$_3$) targets for $^{16}$O beam experiments.
Compared with conventional physical vapor deposition (PVD)\cite{taherisynthesis} techniques such as magnetron sputtering, FCVA provides a high ionization rate, rapid low-temperature deposition, and strong film adhesion. Moreover, the curved magnetic filter in the FCVA system can effectively remove macroparticles generated during the arc discharge process, resulting in dense films with improved surface quality.
The Cr$_2$O$_3$ target has a relatively high oxygen content and good thermal and mechanical stability, and has been used as an oxide target material in oxygen-induced nuclear reaction studies\cite{bromley1959mechanisms, husinsky1986laser, boujlaidi2015photon}. Nevertheless, the actual oxygen areal density and stoichiometric ratio may deviate from the nominal values owing to differences in the preparation conditions. In addition, trace carbon contamination introduced during target fabrication or handling may produce observable beam-induced background signals during in-beam measurements. Therefore, detailed characterization of Cr$_2$O$_3$ targets is required before their application in $^{16}$O+$^{16}$O fusion experiments.

In this work, we systematically study the suitability of Cr$_2$O$_3$ targets prepared by FCVA for low-energy $^{16}$O+$^{16}$O fusion experiments. EBS measurements\cite{mayer1999simnra} were used to determine the elemental composition and areal density of the targets. In addition, in-beam $\gamma$-ray spectra measured under $^{16}$O beam irradiation were analyzed to study the background contributions from impurities, with a focus on reactions related to carbon. Sec.~\ref{sec:target} describes the FCVA preparation of the Cr$_2$O$_3$ targets. Sec.~\ref{sec:exp} presents the EBS measurement and the determination of the chemical stoichiometry and target thickness. Sec.~\ref{sec:result} reports the in-beam $\gamma$-ray measurement and the analysis of impurity-related background contributions. The conclusions are given in Sec.~\ref{sec:summary}.

\section{Target fabrication}
\label{sec:target}  
The Cr$_2$O$_3$ targets used in this study were fabricated using the FCVA technique, which allows precise control of film thickness while minimizing the incorporation of macroparticles.  
Previous applications of FCVA for preparing Ti$^{12}$C\cite{wang2022development} thick targets and Ti$^{18}$O\cite{wangMeasurement18OaG22Ne2023} thin targets have demonstrated excellent irradiation resistance. These results motivated the extension of FCVA to the preparation of Cr$_2$O$_3$ targets.

\begin{figure}[htbp]
		\centering
		\includegraphics[width=8cm]{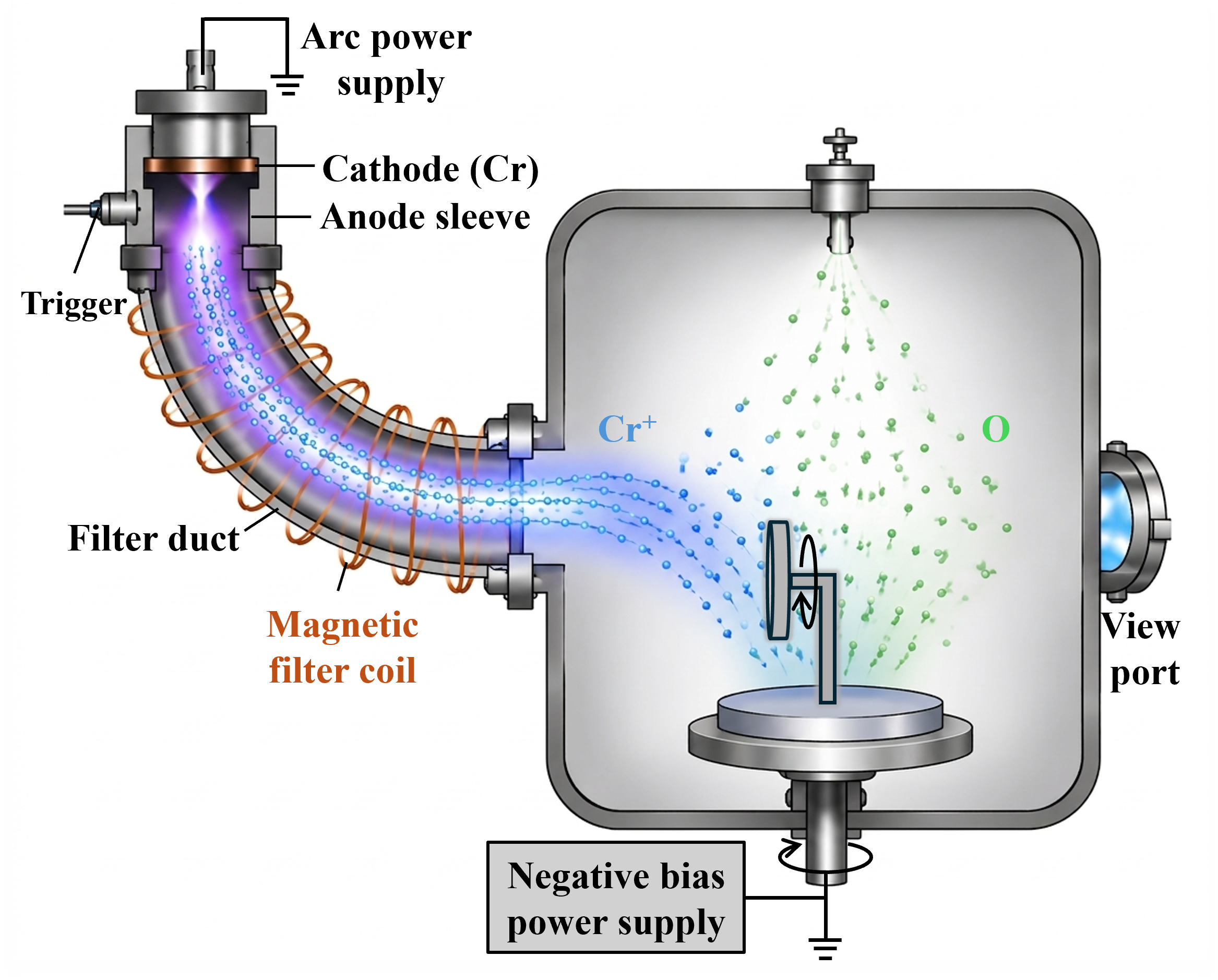}
		\caption{Schematic diagram of the FCVA system.
        }
		\label{fig:FCVA}
	\end{figure}

In this work, the Cr$_2$O$_3$ targets were fabricated using the FCVA deposition system at Beijing Normal University\cite{lin2026developmenta,chen2025enhanced}. A schematic diagram of the FCVA setup is shown in \Fig\ref{fig:FCVA}. The system consists of a high-purity Cr cathode, an anode sleeve, a magnetic filter with copper coils, a vacuum chamber, and a sample stage capable of rotation in both horizontal and vertical directions. During deposition, Cr ions generated by cathodic arc discharge are guided along the axis of a 90$^\circ$ filter duct into the vacuum chamber (pressure was $4 \times 10^{-4}$ Pa), while macroparticles and neutral species are removed through collisions within the filter, ensuring dense and uniform film deposition.
The substrates were Cr disks with a diameter of 15~mm and thickness of 1~mm.
Prior to deposition, the substrates were mechanically polished sequentially using 800, 1200, and 2000 grit wet abrasive papers to remove surface oxides and contaminants, followed by ultrasonic cleaning in anhydrous ethanol for 5~minutes. Before FCVA deposition, the substrate surfaces were further cleaned by Cr-ion sputtering under sequential negative biases of $-800$~V, $-600$~V, and $-400$~V for 20~s each.
Cr was selected as the substrate because its good material compatibility with Cr$_2$O$_3$ is expected to promote strong interfacial bonding.

A metallic Cr transition layer of approximately 1~$\mu$m was first deposited under a $-110$~V bias for 60~min to cover residual surface oxides and contaminants, provide a fresh Cr surface, and improve the adhesion of the subsequent Cr$_2$O$_3$ film.
Oxygen gas was then introduced at 20~sccm (standard cubic centimeter per minute) to co-deposit the Cr$_2$O$_3$ layer for 3~minutes. During deposition, 2~sccm of Ar was continuously supplied to promote cooling and prevent excessive oxidation, thereby avoiding arcing. Using this procedure, two thin targets for $^{16}$O fusion experiments (sample 1 and sample 2) were fabricated. Under the same conditions, Cr$_2$O$_3$ films were also deposited on silicon substrates as representative samples for composition and structural analysis (sample 3). 
\begin{table*}[t]
\renewcommand{\arraystretch}{1.2}
\begin{threeparttable}
\caption{\label{tab:target}
    Oxygen, carbon areal density and Cr/O ratio in the Cr$_2$O$_3$ layer for the targets prepared by FCVA technique in this work.
    }
\begin{tabular*}{\textwidth}{@{\extracolsep{\fill}} lccccc}
\hline
\textbf{Target} & 
\textbf{Layer structure} & 
\makecell{\textbf{$^{16}$O deposition}\\
\textbf{time (min)}} & 
\makecell{\textbf{$^{16}$O areal density\tnote{a}}\\
\textbf{(10$^{17}$ atoms/cm$^{2}$)}} & 
\makecell{\textbf{$^{12}$C areal density\tnote{a}}\\
\textbf{(10$^{15}$ atoms/cm$^{2}$)}} & 
\makecell{\textbf{Cr/O ratio\tnote{a}}}\\
\hline

Sample 1 & Cr\(_2\)O\(_3\) + Cr + Cr & 3 & 
3.24 \(\pm\) 0.24 & 
6.75 \(\pm\) 0.49 & 
0.60 \(\pm\) 0.06 \\

Sample 2 & Cr\(_2\)O\(_3\) + Cr + Cr & 3 & 
3.25 \(\pm\) 0.24 & 
6.60 \(\pm\) 0.48 & 
0.58 \(\pm\) 0.06 \\

Sample 3-1 & Cr\(_2\)O\(_3\) + Cr + Si & 3 & 
3.25 \(\pm\) 0.24 & 
6.44 \(\pm\) 0.47 & 
0.56 \(\pm\) 0.06 \\

Sample 3-2 & Cr\(_2\)O\(_3\) + Cr + Si & 3 & 
3.25 \(\pm\) 0.24 & 
6.45 \(\pm\) 0.47 & 
0.57 \(\pm\) 0.06 \\

\hline
\end{tabular*}

\begin{tablenotes}
\footnotesize
\item[a] Obtained by fitting the EBS spectrum.
\end{tablenotes}

\end{threeparttable}
\end{table*}
    
\section{EBS characterization}
\label{sec:exp}    
The EBS experiments were performed at the 3 MV tandetron accelerator facility of Sichuan University (SCU)\cite{xieTerminalVoltageCalibration2025,han2018ion}. The accelerator is capable of delivering stable light and heavy ion beams with terminal voltages up to 3 MV, providing the beam energies required for the EBS measurements.
It is equipped with a dual ion-source injector, consisting of a duoplasmatron (358) source and an 860A Cs sputter source. The 358 source provides H$^-$ beams up to $\sim$80~$\mu$A and He$^-$ beams up to $\sim$10~$\mu$A, while the 860A Cs source delivers a variety of heavy ion beams, including C$^{-}$, O$^{-}$, Si$^{-}$, Au$^{-}$, and W$^{-}$. The beam transmission efficiency of the system is typically 50--80\%. In EBS, the elemental composition and thickness of the sample are inferred from the energy spectrum of the scattered ions, which directly depends on the incident beam energy. Therefore, the accuracy of the beam energy is crucial.
To improve the reliability of the terminal voltage, Xie \textit{et al.}\cite{xieTerminalVoltageCalibration2025} recently performed a systematic calibration of the tandetron accelerator terminal voltage in the range of 500--2100 kV using several threshold reactions and narrow resonance peaks. 
	\begin{figure}[htbp]
		\centering
		\includegraphics[width=8cm]{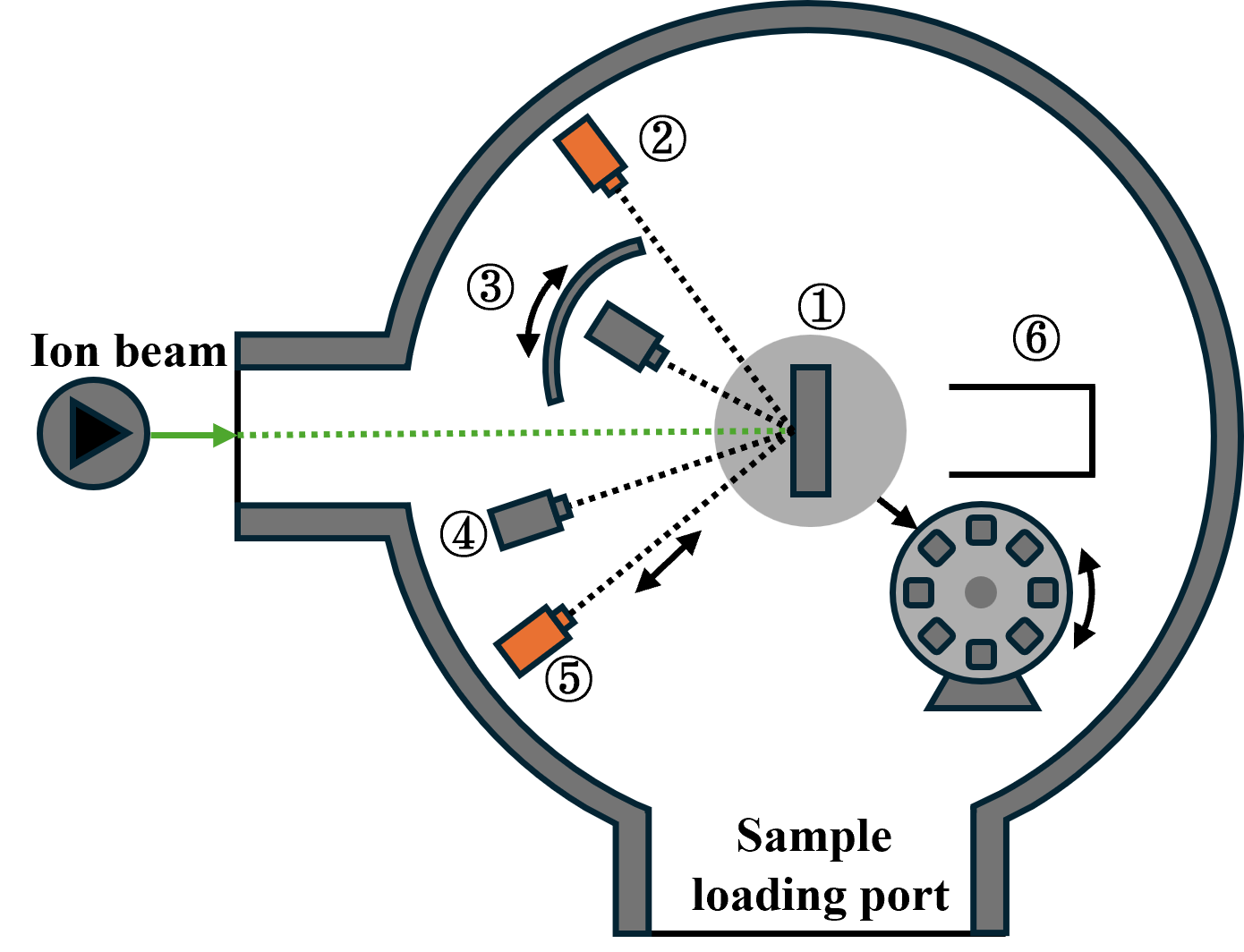}
		\caption{Schematic layout of the IBA vacuum chamber. \ding{172}: Multi-sample stage manipulator. \ding{173}：SDD detector. \ding{174}：Rotatable silicon detector. \ding{175}：Fixed silicon detector. \ding{176}：HPGe detector. \ding{177}：Faraday cup.
        }
		\label{fig:EBS}
	\end{figure}

The EBS experiment was conducted in the ion beam analysis (IBA) terminal, the schematic layout of which is shown in \Fig\ref{fig:EBS}. Two adjustable slits are used for beam collimation, allowing the beam spot size at the target position to be reduced to around 1~mm (FWHM).
A multi-sample target holder accommodating up to 25 samples is equipped with a three-dimensional motorized positioning system for precise rotation and translation under vacuum. The rotational and translational accuracies are 0.01$^\circ$ and 0.01~mm, respectively. A Faraday cup is placed behind the holder for charge collection, and a 300V positive bias is applied to the target stage for accurate charge integration in thick-target measurements.
EBS measurements were performed using two ion-implanted silicon detectors (U-012-050-100, AMETEK ORTEC, Oak Ridge, TN, USA). One detector is angle-tunable from $-180^\circ$ to $180^\circ$ (excluding $10^\circ$--$50^\circ$) and located at $\sim$10~cm from the target. The other is fixed at 170$^\circ$ with a distance of $\sim$18~cm and was used in this work, with a 100~$\mu$m depletion layer and a 50~mm$^2$ active area.
An HPGe detector (GL0055, Mirion Technologies, Meriden, CT, USA) and an SDD X-ray detector (XR-100CR, Amptek Inc., Bedford, MA, USA) are installed at 135$^\circ$ relative to the beam direction. The distance between HPGe detector and the target is adjustable, while the SDD detector is fixed at $\sim$21~cm from the target center. To avoid resonance regions in the $^{16}$O scattering cross section while maintaining sufficient penetration depth in the Cr$_2$O$_3$ layer, a proton beam energy of 2.488~MeV was selected. The proton beam was incident normal to the target surface. The beam current was continuously monitored and kept below 2~nA to minimize pile-up effects, resulting in a dead time of less than 2\%. To assess target homogeneity, the sample with Si substrate was additionally measured at a position 2~mm away from the target center.

\begin{figure}[htbp]
		\centering
		\includegraphics[width=8cm]{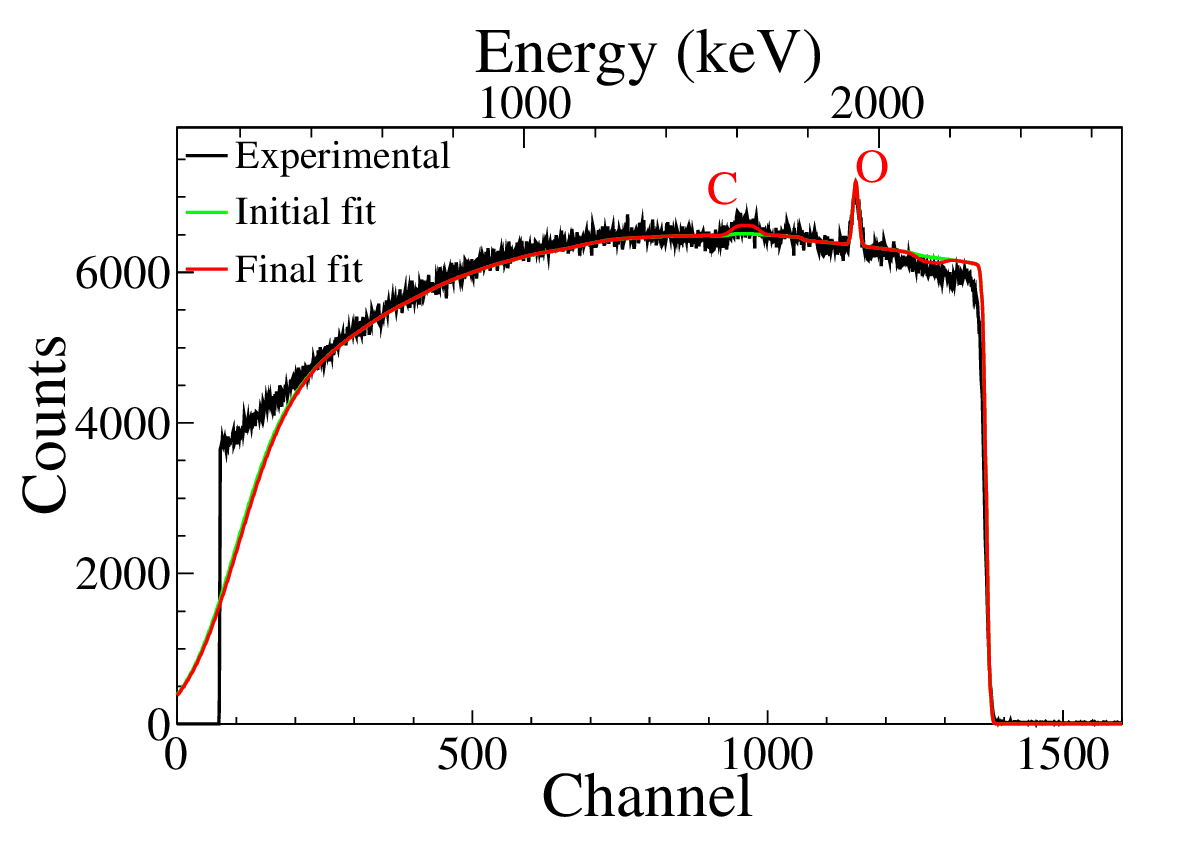}
		\caption{Comparison of initial and optimized SIMNRA fits for the Sample 1 EBS spectrum.
        }
		\label{fig:compare-fit}
	\end{figure}

\begin{figure*}[htbp]
		\centering
		\includegraphics[width=\textwidth]{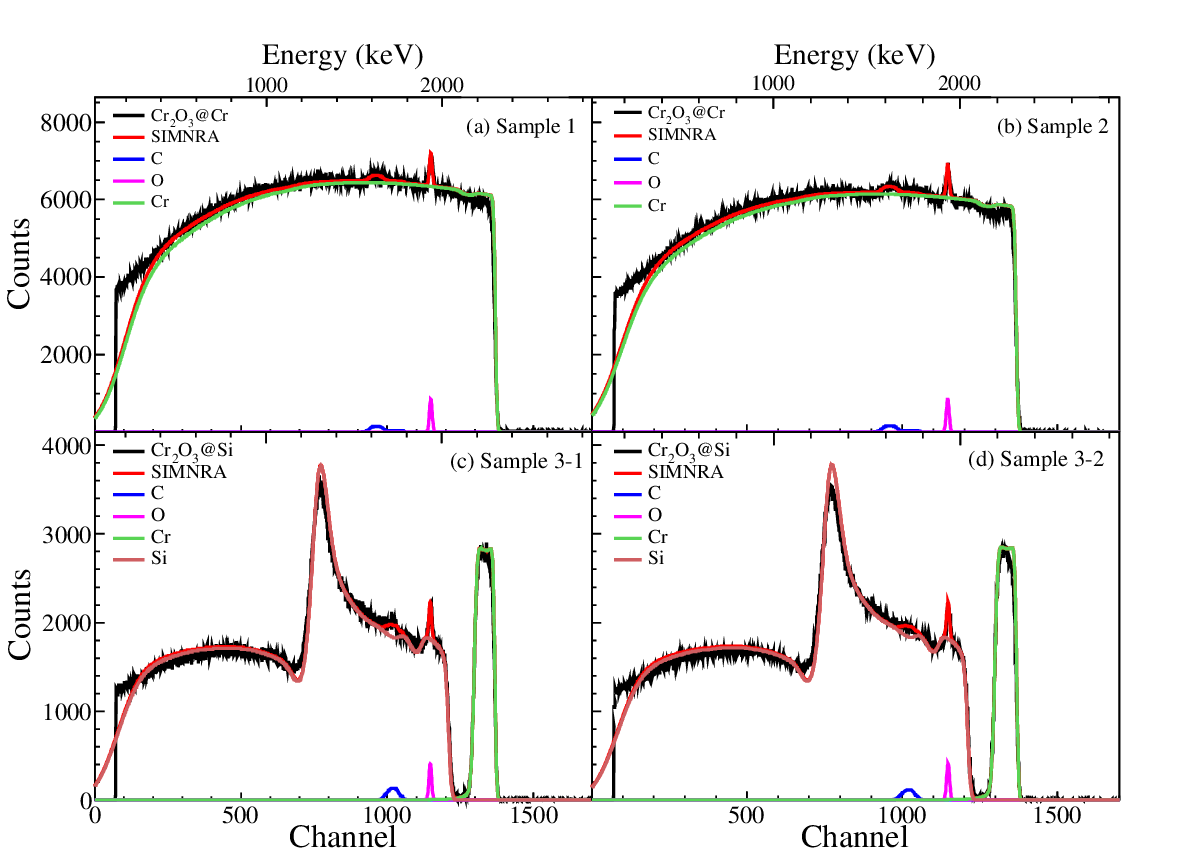}
		\caption{Experimental EBS spectra and the fitting results from SIMNRA. The spectra of sample 3-2 are scaled by a factor of 0.45 for better display.
        }
		\label{fig:EBS-result}
	\end{figure*}

To quantitatively characterize the Cr$_2$O$_3$ targets, the experimental EBS spectra were analyzed using the SIMNRA program\cite{mayersimnra,liu2021tritium,chen2020bixs,fu2014tritium, wu2013pixerbs}. In the analysis, SRIM stopping powers\cite{SRIM}, Chu and Yang energy-loss straggling models\cite{chu1976calculationa, yang1991empirical}, Andersen screened Rutherford cross sections\cite{andersen1980largeangle} were used. Both multiple and dual scattering were considered, with multiple scattering modeled by Amsel's approach\cite{amsel2003small}. The cross section data for $^{16}$O($p$, $p_{0}$)$^{16}$O, $^{12}$C($p$, $p_{0}$)$^{12}$C, and $^{nat}$Si($p$, $p_{0}$)$^{nat}$Si were taken from Ref.~\cite{amirikas1993measurement}, corresponding to approximately 5.1\% uncertainty, while $^{nat}$Cr($p$, $p_{0}$)$^{nat}$Cr was obtained from evaluations in the SigmaCalc archive\cite{gurbich2016sigmacalc}.
Based on the FCVA deposition process, the target was modeled as a three-layer structure consisting of a Cr$_2$O$_3$ surface layer, an intermediate Cr layer, and a semi-infinite Cr/Si substrate. The thickness of the Cr$_2$O$_3$ layer was initially estimated from the deposition time and growth rate, corresponding to $3.0\times10^{17}$ atoms/cm$^2$, while the intermediate Cr layer was set to $6.0\times10^{18}$ atoms/cm$^2$. 
For the initial SIMNRA input, a stoichiometric Cr$_2$O$_3$ layer (Cr:O = 2:3) was assumed, while a small uniform carbon content was introduced in both the oxide layer and the substrate to account for possible contamination during deposition and handling.
Fig.~\ref{fig:compare-fit} shows the initial and final fits of the EBS spectrum of the sample 1. Although the main Cr and O signals were well reproduced, clear discrepancies were observed in the 1500\textendash1700 keV region, indicating that the assumption of a uniform carbon distribution was insufficient.
By iteratively adjusting the carbon depth profile in the SIMNRA fit, a significantly improved agreement between the calculated and experimental spectra was obtained, particularly for the spectral features associated with carbon, as illustrated by the final fit in Fig.~\ref{fig:compare-fit}. The experimental EBS spectra and the final SIMNRA fits for all samples are shown in Fig.~\ref{fig:EBS-result}. For clarity and to allow a unified comparison, the experimental and SIMNRA spectra of Sample~3-2 were rescaled by a constant factor of 0.45. The extracted oxygen areal density, Cr/O atomic ratio, and carbon impurity content are summarized in Table~\ref{tab:target}. Samples~1 and 2 correspond to Cr$_2$O$_3$ films deposited on Cr substrates, while Samples~3 were prepared on Si substrates under identical FCVA conditions to evaluate film uniformity. The Si substrates were selected to reduce the interference from the thick Cr background in EBS measurements, allowing a more reliable assessment of the spatial uniformity of the deposited Cr$_2$O$_3$ films. The spectra obtained at the beam center and the off-center position are denoted as Sample 3-1 and Sample 3-2, respectively. 
    
The $^{16}$O areal densities of the targets are in the range of 
$(3.24$--$3.25)\times10^{17}$~atoms/cm$^2$, indicating excellent thickness uniformity and stable deposition control. The uncertainty in oxygen content is estimated at 7.3\%, including 5.1\% from scattering cross section, less than 5\% from SRIM stopping power, 1\% SIMNRA fitting error and 0.7\% statistical error. The Cr/O atomic ratios are close to the stoichiometric value of Cr$_2$O$_3$, although slightly lower than the ideal value of 0.667, indicating a small oxygen enrichment. The uncertainty of the Cr/O ratio was obtained by propagating the uncertainties of the measured Cr and O areal densities. The slight oxygen enrichment does not necessarily indicate the formation of a distinct CrO$_3$ phase, since EBS determines the overall elemental composition but cannot directly distinguish between different chromium oxide phases. In the FCVA process, the oxygen content is affected by the oxygen flow rate and corresponding oxygen partial pressure, while the substrate temperature was not independently monitored or controlled. The observed deviation from the ideal Cr$_2$O$_3$ stoichiometry may therefore be related to the oxidation conditions during deposition, such as the Ar-to-O ratio, and is consistent with the observations reported by Silva\cite{silvaComparativeAnalysisAnodized2014} \textit{et al.} for Ta$_2$O$_5$ targets. The agreement between Samples~3-1 and 3-2 further confirms good spatial uniformity within the effective beam irradiation area.
The slight difference in the high energy edge of Cr scattering peak between Samples 3-1 and 3-2 maybe attributed to the sample surface roughness.

The $^{12}$C areal densities are in the range of 
$(6.44$--$6.75)\times10^{15}$~atoms/cm$^2$, corresponding to a carbon atomic fraction of approximately 1.25\textendash1.29\% in the Cr$_2$O$_3$ layer. The uncertainty of the $^{12}$C areal density is 7.3\%, identical to that of the $^{16}$O areal density due to the same EBS fitting procedure and uncertainty sources. However, a large amount of $^{12}$C contamination, with an areal density of $2.55\times10^{17}$~atoms/cm$^2$, is observed on the surface of the Cr substrate, which is about 40 times higher than that in the Cr$_2$O$_3$ layer. 
The mechanical polishing and the plasma cleaning procedures seem unable to fully remove the carbon contaminants from the Cr substrate.
This substantial carbon contamination in the Cr substrate can enhance the characteristic $\gamma$-ray yields from $^{24}$Mg and $^{27}$Al, thereby affecting the determination of the $^{16}$O+$^{16}$O fusion cross section (see next section).
In addition，the SIMNRA fits also show a slightly higher C content in the intermediate Cr layers of Samples~3-1 and 3-2. This may be related to the simultaneous Cr-ion plasma cleaning of the Cr and Si substrates, during which carbon-containing surface contaminants on the Cr substrates could have been sputtered and redeposited onto the Si substrates.

\section{In-beam $\gamma$-ray Spectroscopy and Impurity Analysis}
\label{sec:result}
	\begin{figure}[htbp]
		\centering
		\includegraphics[width=8cm]{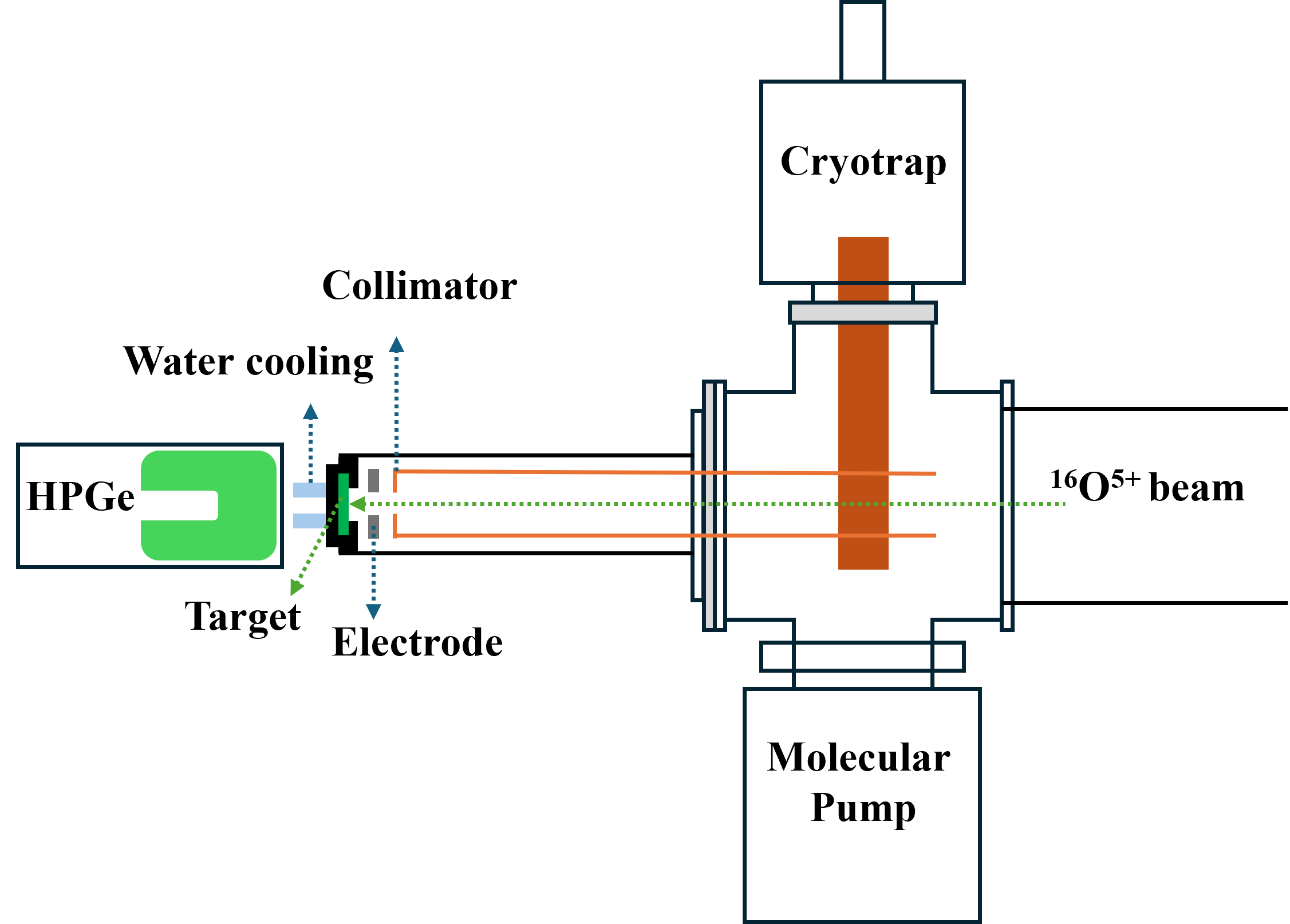}
		\caption{Schematic diagram of the experimental setup for in-beam $\gamma$-ray measurement.
        }
		\label{fig:exp}
	\end{figure}

	\begin{figure}[htbp]
		\centering
		\includegraphics[width=8cm]{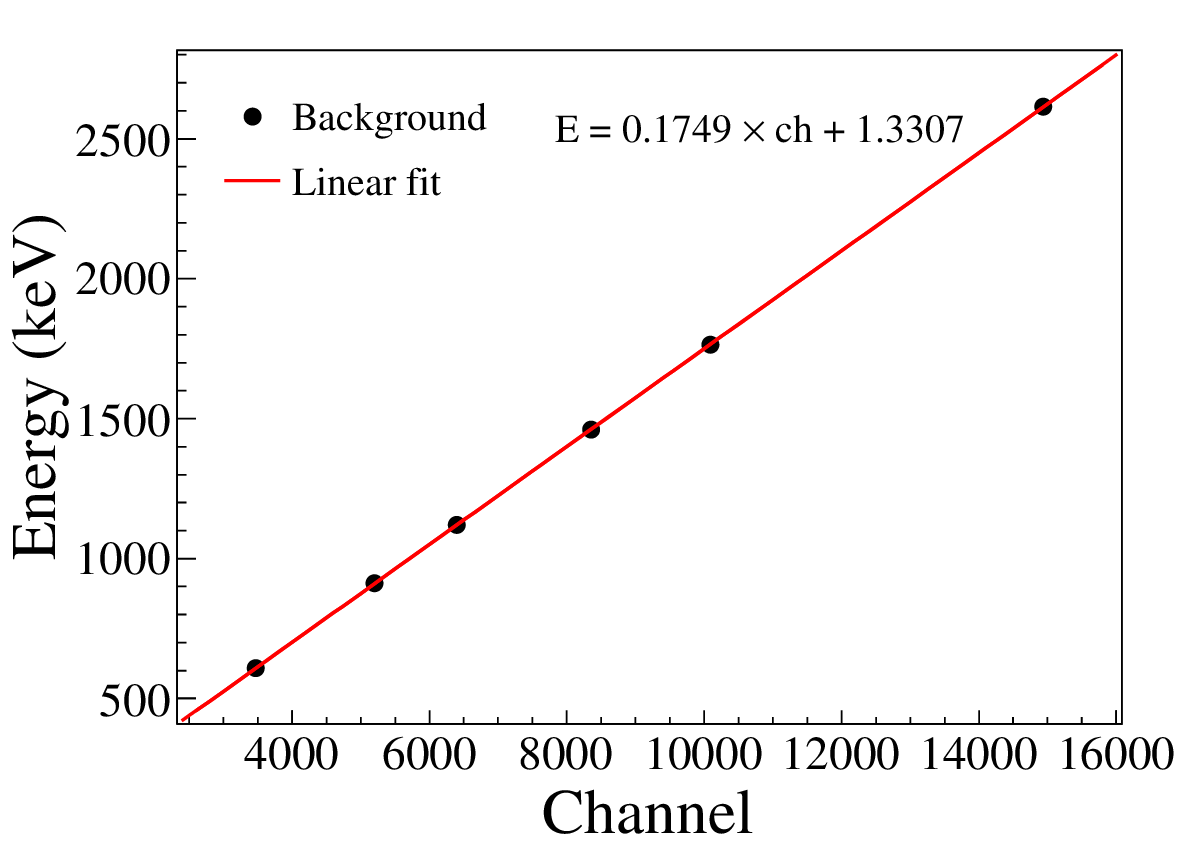}
		\caption{Energy calibration of the HPGe detector using the $\gamma$ rays from environmental background.
        }
		\label{fig:energy-cal}
	\end{figure}

	\begin{figure*}[htbp]
		\centering
        \includegraphics[width=\textwidth]{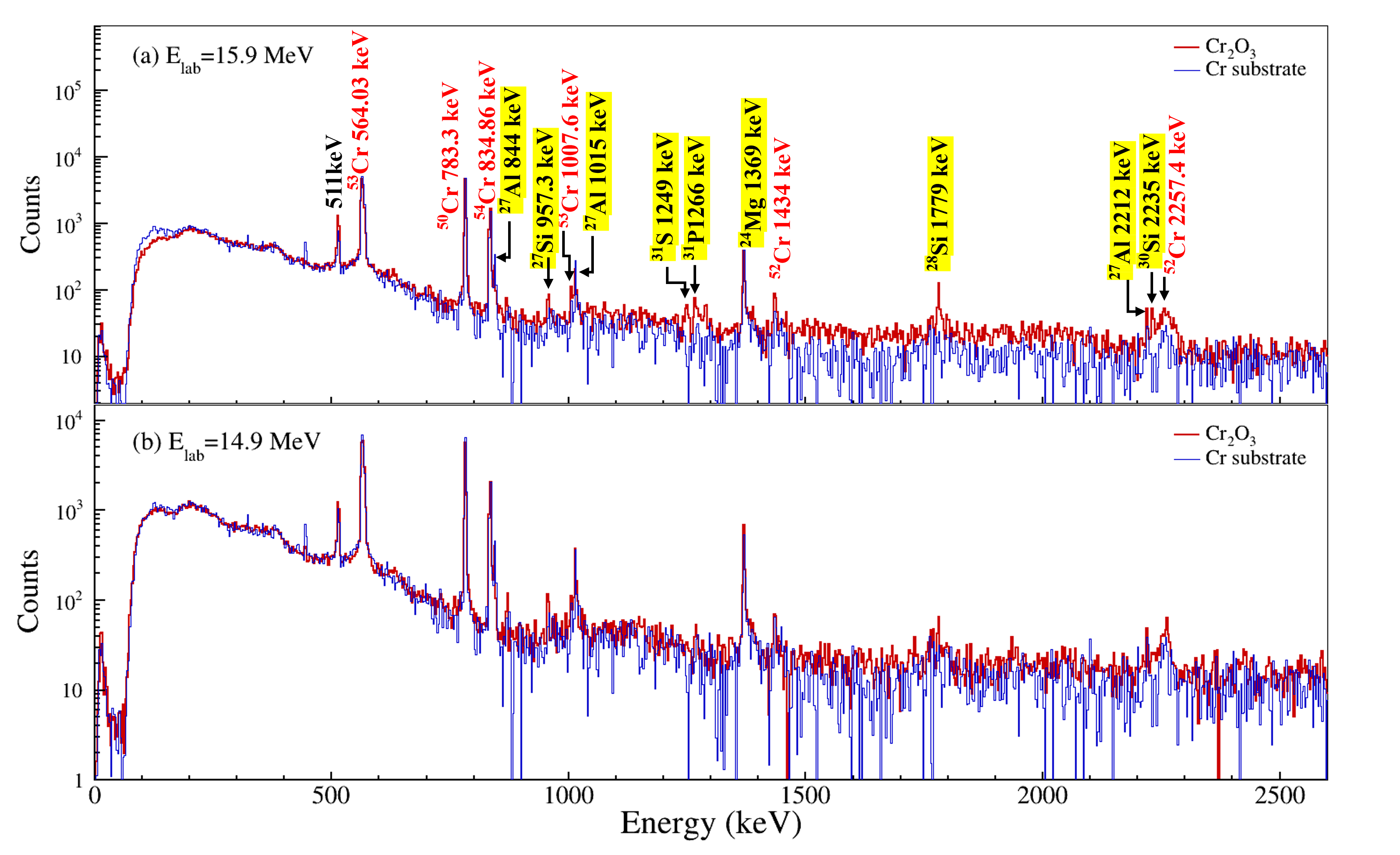}
        \caption{Background-subtracted in-beam $\gamma$-ray spectra measured by the HPGe detector under $^{16}$O$^{5+}$ beam irradiation at (a) $E_{\mathrm{lab}} = 15.9$ MeV and (b) $E_{\mathrm{lab}} = 14.9$ MeV. The spectra at $E_{\mathrm{lab}} = 15.9$ MeV were normalized using the 783.3 keV $\gamma$-ray peak from Cr Coulomb excitation, while the Cr substrate spectrum at $E_{\mathrm{lab}} = 14.9$ MeV was normalized to the beam current. The characteristic $\gamma$-ray energies and their origins are indicated.}
		\label{fig:16O-gamma}
	\end{figure*}       
The in-beam $\gamma$-ray measurement was carried out at the Nuclear Physics Experiment (NPE) terminal of the 3 MV tandetron accelerator at SCU. An O$^{5+}$ beam was used to bombard the Cr$_2$O$_3$ target. A schematic diagram of the experimental setup is shown in \Fig\ref{fig:exp}.
A liquid-nitrogen-cooled copper cold trap was installed upstream of the beam line to suppress carbon deposit and improve the vacuum conditions in the target chamber. A Ta collimator coated with a high-purity Cr layer ($>$1~$\mu$m) was fixed at the end of the cold trap to limit the beam spot. A suppression ring biased at $-600$~V was positioned approximately 1~cm downstream of the collimator and 5~cm upstream of the target to suppress secondary electrons emitted from the target surface.
The reaction target was mounted at the end of the vacuum pipe and cooled by a circulating deionized water system to avoid thermal damage and leakage currents. The target pipe and the cold trap section were electrically insulated from each other. During the experiment, the vacuum chamber pressure was maintained around $1\times10^{-5}$~Pa.
The in-beam $\gamma$-ray spectra were measured using a 50\% relative-efficiency coaxial low-energy N-type HPGe detector (GMX50-83-CW，AMETEK ORTEC, USA) positioned 4 cm downstream of the target. The detector employed a standard coaxial geometry with a crystal diameter of 62~mm and length of 85.3~mm, while the central bore had a diameter and depth of 9.2~mm and 76.4~mm, respectively. 
A 0.9~mm thick carbon-fiber entrance window was used to improve the transmission efficiency for low-energy $\gamma$ rays. The detector exhibited an energy resolution of 1.9~keV at the 1.33~MeV $\gamma$ ray of $^{60}$Co and 866~eV at the 122~keV $\gamma$ ray of $^{152}$Eu, demonstrating excellent energy resolution performance. 
Energy calibration was performed using environmental background $\gamma$ rays from $^{40}$K ($1460.8$ keV), $^{208}$Tl ($2614.5$ keV), $^{214}$Bi ($609.3$, $1120.3$, and $1764.5$ keV), and $^{228}$Ac ($911.2$ keV), as shown in \Fig\ref{fig:energy-cal}.
To investigate impurity components in the Cr$_2$O$_3$ target and their impact on fusion experiments, in-beam $\gamma$-ray spectra were measured under 15.9 and 14.9 MeV $^{16}$O$^{5+}$ beams bombardment with an intensity about 2 $\mu$A for the Cr$_2$O$_3$ target. The two beam energies were selected to confirm the contribution of the $^{16}$O+$^{16}$O fusion reaction to the characteristic $\gamma$-rays, and the 15.9 MeV beam energy is expected to show an enhancement in the specific $\gamma$-rays. The substrate contribution was estimated by bombarding the 14.9 MeV $^{16}$O$^{5+}$ beam to a Cr substrate, with the same treatment as that used to fabricate the Cr$_2$O$_3$ targets. The environmental background spectra measured without beam irradiation were normalized according to the corresponding acquisition times and subtracted from the beam-on spectra.
Fig.~\ref{fig:16O-gamma} shows the beam-on spectra at incident energies of $E_{\mathrm{lab}} = 15.9$ MeV and $14.9$ MeV in (a) and (b), respectively. The Cr substrate spectrum measured at $E_{\mathrm{lab}} = 14.9$ MeV was normalized to the same beam charge of that in Cr$_2$O$_3$ target in Fig.~\ref{fig:16O-gamma} (b). For comparison, this Cr substrate spectrum measured at $E_{\mathrm{lab}} = 14.9$ MeV is also shown in Fig.~\ref{fig:16O-gamma} (a), where it was normalized to the same yield of the 783.3 keV $\gamma$ peak from Cr Coulomb excitation. The characteristic $\gamma$-ray energies and their corresponding origins are indicated in the figure.

Clear $\gamma$-ray peaks from $^{50,52,53,54}$Cr (e.g., 564.0, 783.3, 834.8, 1007.6, 1289.5, 1434.0, and 2257.4 keV) are observed in both the Cr$_2$O$_3$ target and pure Cr substrate spectra. These peaks are mainly caused by Coulomb excitation of Cr nuclei induced by the $^{16}$O beam.
Several $\gamma$-ray peaks that are unique to the oxygen-containing target, such as those at 1249 keV, 1266 keV, 1779 keV, and 2235 keV, are clearly observed at $E_{\mathrm{lab}} = 15.9$ MeV. These peaks correspond to evaporation channels of the $^{16}$O+$^{16}$O fusion reaction, including $^{16}$O($^{16}$O, $n$)$^{31}$S, $^{16}$O($^{16}$O, $p$)$^{31}$P, $^{16}$O($^{16}$O, $\alpha$)$^{28}$Si, and $^{16}$O($^{16}$O, $2p$)$^{30}$Si, indicating clear fusion signatures at the present beam energy and can be used to extract the cross sections of those channels.
A comparison of the spectra measured at 15.9 and 14.9 MeV shows that the intensity of the 1779 keV transition decreases significantly at the lower beam energy. This behavior is consistent with the rapid decrease of the $^{16}$O+$^{16}$O fusion cross section toward lower energies.
Both the Cr$_2$O$_3$ target and the Cr substrate spectra exhibit pronounced $\gamma$-ray peaks associated with $^{27}$Al and $^{24}$Mg, including 844 keV, 1015 keV, and 1369 keV. These nuclei can be produced not only in $^{16}$O+$^{16}$O evaporation channels but also in $^{12}$C+$^{16}$O fusion reactions induced by carbon impurities. 
This indicates that the $^{27}$Al and $^{24}$Mg peaks are dominated by $^{12}$C+$^{16}$O fusion reactions at the substrate, with a minor contribution from $^{16}$O+$^{16}$O under the present conditions.
The carbon contaminant, especially in the Cr substrate, strongly prevents the extraction of the partial cross sections of the $^{16}$O($^{16}$O, $\alpha$p)$^{27}$Al and $^{16}$O($^{16}$O, $2\alpha$)$^{24}$Mg channels.
Although Doppler effects are generally expected for in-beam $\gamma$-rays due to recoil motion of excited nuclei, no resolvable Doppler shift is observed within the energy resolution of the present setup for all identified transitions.
Combined with the EBS analysis, the comparison between the Cr$_2$O$_3$@Cr target and the pure Cr substrate spectra indicates that the observed beam-induced background is mainly associated with carbon impurities in the target substrate. The presence of strong $^{12}$C+$^{16}$O reaction signatures in the pure Cr substrate spectrum highlights the important contribution of substrate-related carbon contamination. 
The relatively high carbon areal density on the Cr substrate surface may result from contamination during substrate preparation, handling, or storage.
Therefore, cleaner substrate materials should be considered for future low-background $^{16}$O+$^{16}$O fusion measurements.

\section{Conclusions}
\label{sec:summary}
In this work, Cr$_2$O$_3$ solid targets for low-energy $^{16}$O+$^{16}$O fusion experiments were prepared using the FCVA technique. EBS and in-beam $\gamma$-ray spectroscopy measurements were performed to systematically investigate the chemical composition, thickness uniformity, and impurity content of the targets. The EBS results indicate that the deposited Cr$_2$O$_3$ films exhibit stable stoichiometry, good spatial uniformity, and sufficient $^{16}$O areal density, with values of $(3.24$--$3.25)\times10^{17}$~atoms/cm$^2$. The thickness and oxygen content are well controlled and satisfy the requirements for low-energy $^{16}$O+$^{16}$O fusion measurements.
SIMNRA fitting results show that the Cr$_2$O$_3$ oxide layer contains a carbon atomic fraction of approximately 1.25--1.29\%, while additional carbon contamination is also present in the Cr substrate. In-beam $\gamma$-ray spectroscopy further revealed characteristic $\gamma$-ray peaks induced by $^{12}$C+$^{16}$O fusion reactions originating from these carbon impurities, including the 844~keV, 1015~keV, and 1369~keV transitions associated with the de-excitation of $^{27}$Al and $^{24}$Mg.
Four characteristic $\gamma$-ray peaks at 1249, 1266, 1779, and 2235~keV, corresponding to evaporation channels of the $^{16}$O+$^{16}$O fusion reaction, were found to be free from carbon-induced backgrounds and can therefore be used for the extraction of the $^{16}$O+$^{16}$O fusion cross section.

Further improvements for future $^{16}$O+$^{16}$O fusion studies should focus on reducing the carbon contamination in the substrate as well as in the Cr$_2$O$_3$ target to improve the signal-to-background ratio and to allow measuring the total cross sections of the $^{16}$O+$^{16}$O fusion reaction in all channels.
The present work provides an experimental basis for the development of high-purity oxide targets and the optimization of target configurations for future low-background $^{16}$O+$^{16}$O fusion measurements.
\section*{Acknowledgments}
The present work was supported by the National Natural Science Foundation of China (Grant No. 12175156, 12275026, 11805138, 12175158). This work was also supported by Institutional Research Fund from Sichuan University (Grant No. 2023SCUNL102).


\end{document}